\documentclass[reprint,amsmath,amssymb,superscriptaddress,floatfix,prl,aps]{revtex4-2}
\usepackage{graphicx}
\usepackage{color}
\usepackage{hyperref}
\usepackage{xltabular}
\usepackage{makecell}
\usepackage{multirow}
\usepackage{booktabs}
\usepackage{stmaryrd}
\usepackage[T1]{fontenc}

\begin{document}

\title{Tilt-driven ferrielectricity in PbZrO$_3$}

\author{Huazhang Zhang}
\thanks{These authors contributed equally to this work}
\email[Corresponding author: ]{zhanghz782@126.com}
\affiliation{School of Physics and Mechanics, Wuhan University of Technology, Wuhan 430070, China}
\affiliation{Theoretical Materials Physics, Q-MAT, University of Liège, B-4000 Sart-Tilman, Belgium}

\author{Ying Liu}
\thanks{These authors contributed equally to this work.}
\affiliation{School of Aerospace, Mechanical and Mechatronic Engineering, The University of Sydney, Sydney, NSW 2008, Australia}
\affiliation{Catalan Institute of Nanoscience and Nanotechnology (ICN2), Campus Universitat Autonoma de Barcelona, Barcelona 08193, Spain}

\author{Carson Carroll}
\affiliation{School of Physical Sciences, University of Kent, Canterbury CT2 7NH, United Kingdom}

\author{Saptam Ganguly}
\affiliation{Catalan Institute of Nanoscience and Nanotechnology (ICN2), Campus Universitat Autonoma de Barcelona, Barcelona 08193, Spain}

\author{Bin Xu}
\affiliation{Jiangsu Key Laboratory of Frontier Material Physics and Devices, Jiangsu Key Laboratory of Advanced Negative Carbon Technologies, School of Physical Science and Technology, Soochow University, Suzhou 215006, China}

\author{Xiaozhou Liao}
\affiliation{School of Aerospace, Mechanical and Mechatronic Engineering, The University of Sydney, Sydney, NSW 2008, Australia}

\author{Gustau Catalan}
\affiliation{Catalan Institute of Nanoscience and Nanotechnology (ICN2), Campus Universitat Autonoma de Barcelona, Barcelona 08193, Spain}
\affiliation{Institut Català de Recerca i Estudis Avançats (ICREA), Barcelona 08010, Catalunya}

\author{Philippe Ghosez}
\affiliation{Theoretical Materials Physics, Q-MAT, University of Liège, B-4000 Sart-Tilman, Belgium}

\author{Nicholas C. Bristowe}
\email[Corresponding author: ]{nicholas.bristowe@durham.ac.uk}
\affiliation{Centre for Materials Physics, Durham University, Durham DH1 3LE, United Kingdom}

\date{\today}

\begin{abstract}
We reveal a tilt-driven mechanism for ferrielectricity in prototypical antiferroelectric PbZrO$_3$. 
Specifically, introducing an additional octahedral tilt into the antiferroelectric $Pbam$ phase breaks the symmetry constraint that enforces equal antiparallel dipoles, converting the compensated ``$\uparrow \uparrow \downarrow \downarrow$'' nonpolar configuration into an uncompensated ``$\uparrow \uparrow \shortdownarrow \shortdownarrow$'' polar $Pmc2_1$ phase. 
First-principles calculations show that the $Pmc2_1$ phase becomes stabilized under lattice contraction and gains increasing free-energy advantage over competing phases at finite temperatures.
Atomic-scale imaging directly confirms the presence of $Pmc2_1$-like structures in thin films and single crystals. 
This work identifies a symmetry-governed, kinetically easily accessible pathway to ferrielectricity and establishes a form of ``competitive'' improper ferroelectricity, with broad implications for antiferroelectrics.
\end{abstract}

\maketitle

\newpage

Antiferroelectrics (AFEs) are typically characterized by antiparallel dipoles that can be aligned parallel under an applied electric field \cite{KittelAFE, RabeAFE, LinesGlass, gustau2025}.
PbZrO$_3$ (PZO), the prototypical AFE, exhibits a ``$\uparrow \uparrow \downarrow \downarrow$'' antipolar ordering in its $Pbam$ phase \cite{RN534, RN93, RN74}.
Ideally, AFEs show zero remanent polarization in double $P$-$E$ loops \cite{RN1279}; however, real PZO systems often exhibit finite remanent polarization \cite{RN697, RN505, RN702, RN98}, which has been suggested to arise from possible ferroelectric or ferrielectric (FiE) phases \cite{RN98, RN90, RN853, RN824, RN1280}. 

FiE phases in PZO have recently attracted significant attention, with several polar states exhibiting antiparallel but uncompensated dipole arrangements reported theoretically and experimentally.
Atomic-resolution scanning transmission electron microscopy (STEM) reveals modulated FiE structures in chemically modified PZO \cite{RN418, RN922}.
For pure PZO, first-principles calculations predict a ``$\uparrow \uparrow \downarrow$'' FiE ground state \cite{RN90}, which has subsequently been observed experimentally \cite{RN933, RN999}.
In addition, in-situ X-ray diffraction identifies an eightfold periodic ``$\uparrow \uparrow \uparrow \uparrow \downarrow \uparrow \uparrow \downarrow$'' FiE configuration under electric field \cite{RN717}.
Notably, these FiE structures are not symmetry-connected to the AFE $Pbam$ phase; their formation requires a drastic reconstruction of the dipole ordering, introducing substantial kinetic barriers that confine them to domain boundaries or transient regions \cite{RN853, RN919}.

In this study, we reveal a tilt-driven FiE mechanism in PZO. 
Introducing additional oxygen octahedral in-phase tilts into the $Pbam$ phase lifts the symmetry constraint that enforces equal magnitude of antiparallel dipoles, transforming the fully compensated ``$\uparrow \uparrow \downarrow \downarrow$'' AFE configuration into an uncompensated ``$\uparrow \uparrow \shortdownarrow \shortdownarrow$'' FiE state with $Pmc2_1$ symmetry.
Unlike the previously reported FiE configurations, this $Pmc2_1$ phase is kinetically favorable as it can be reached from $Pbam$ by adjusting dipole magnitudes without dipole reversal.
First-principles calculations show that the $Pmc2_1$ phase can be stabilized under lattice-contraction conditions, such as hydrostatic pressure and epitaxial strain, which promote the octahedral tilts that drive the mechanism.
Furthermore, aberration-corrected scanning transmission electron microscopy directly reveals the $Pmc2_1$-like FiE structures in both PZO thin films and single crystals. 

Our findings not only uncover a hidden FiE phase in PZO, advancing the understanding of the observed finite remanent polarization, but also point to a special form of ``competitive'' improper ferroelectricity, where macroscopic polarization is governed by subtle competition among multiple improper couplings, extending beyond the standard trilinear coupling and hybrid-improper mechanism \cite{Bousquet2008, Benedek2011}.
The tilt-driven FiE mechanism is not material-specific but instead relies on the coexistence of polarization modulation and in-phase octahedral tilts, as supported by examples beyond PZO.
These results establish a symmetry-based route to FiE and provide a basis for understanding and tuning polarization from competing lattice-mode couplings.

\begin{figure}[t!]
\centering
\includegraphics[scale=0.50]{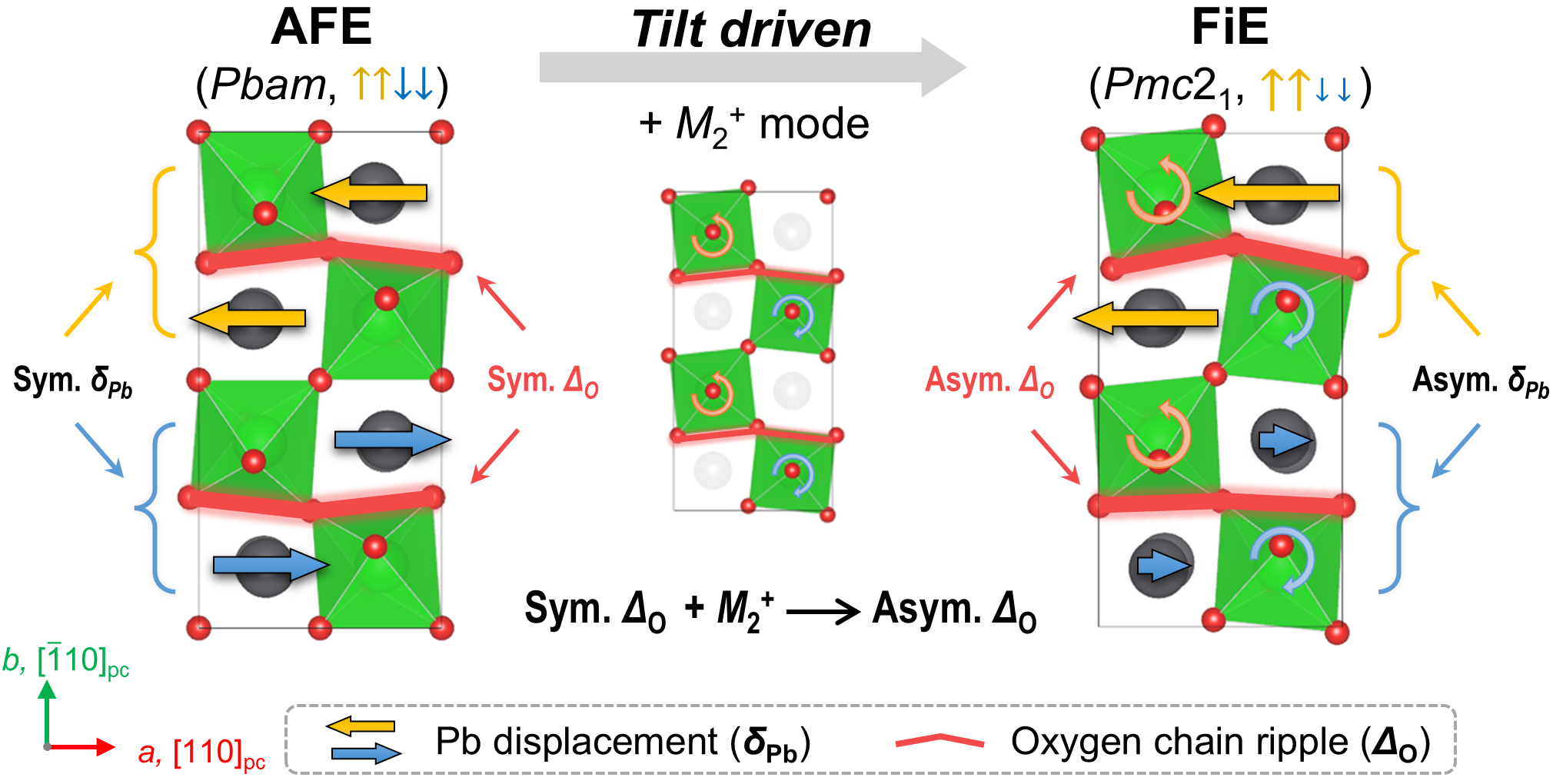}
\caption{Symmetry-based mechanism of tilt-driven FiE in PZO.
The AFE $Pbam$ phase features antiparallel ``$\uparrow \uparrow \downarrow \downarrow$'' Pb displacements and symmetric oxygen-chain ripples. 
The superposition of an in-phase octahedral tilt $M_2^+$ mode renders these ripples to become asymmetric, thereby lifting the symmetry constraints and leading to an uncompensated ``$\uparrow \uparrow \shortdownarrow \shortdownarrow$'' FiE configuration.
}
\label{fig:symmetry}
\end{figure}

We begin by explaining the symmetry-based mechanism through which the octahedral tilt drives FiE in PZO.
As shown in Fig. \ref{fig:symmetry}, in the AFE $Pbam$ phase, the ``$\uparrow \uparrow \downarrow \downarrow$'' polarization pattern originates primarily from the Pb displacements, which belong to the $\Sigma_2$ mode (Fig. S1 \cite{SM}) of the cubic reference structure.
This mode simultaneously involves associated oxygen motions, which give rise to rippling of oxygen chains, as highlighted by the red zigzag lines.
Within $Pbam$, the opposite Pb displacements are symmetry-related and constrained to have equal magnitudes (denoted as ``Sym. $\delta_{\rm Pb}$'').
Accordingly, the oxygen-chain rippling between the ``$\uparrow \uparrow$'' Pb pairs is also symmetric to that between the ``$\downarrow \downarrow$'' Pb pairs (marked as ``Sym. $\Delta_{\rm O}$'').

When an additional in-phase octahedral tilt ($M_2^+$ mode, Fig. S1 \cite{SM}) is imposed on the $Pbam$ structure, the symmetry constraint is broken. 
\nocite{abinit2025, PBEsol, Conserving, RN1102, PseudoDojo, isodistort, INVARIANTS, DFPT, QHA, RN90, HYTCH1998131, RN1157, RN821, RN77, RN1282, RN814, RN1255, RN600}
As illustrated in Fig. \ref{fig:symmetry}, the octahedral tilt superimposes additional distortions onto the oxygen chains, rendering the rippling of the two oxygen chains asymmetric (``Asym. $\Delta_{\rm O}$''). 
This asymmetry creates distinct local environments for the two antiparallel Pb sublattices, thereby lifting the symmetry restriction that dictated equal Pb displacement magnitudes.
Consequently, unequal Pb displacements emerge (``Asym. $\delta_{\rm Pb}$''), producing an uncompensated ``$\uparrow \uparrow \shortdownarrow \shortdownarrow$'' FiE configuration. 

The realization of this mechanism hinges on stabilizing the additional tilt mode on top of the $Pbam$ structure. 
Given the corner-sharing connectivity of the perovskite octahedral framework, lattice contraction generally favors tilts.
Therefore, we carried out first-principles relaxations under hydrostatic pressure, starting from $Pbam$ with an imposed in-phase tilt, to examine whether a FiE ``$\uparrow \uparrow \shortdownarrow \shortdownarrow$'' configuration can be stabilized.
Calculation details are provided in Supplementary Section S1.1 \cite{SM}.

Our structural relaxations converge to a polar phase with space group $Pmc2_1$ under sufficient hydrostatic pressure.
Notably, this phase exhibits ``$\uparrow \uparrow \shortdownarrow \shortdownarrow$'' FiE ordering (Fig. S2 \cite{SM}), yielding a net polarization (Fig. S3 \cite{SM}).
The minimum pressure required to stabilize the $Pmc2_1$ phase is about 5 GPa, corresponding to an average linear strain of $-1.2\%$ relative to the zero-pressure $Pbam$ phase.
Taking the $Pmc2_1$ structure at 10 GPa as an example (structural parameters listed in Table S1 \cite{SM}), it exhibits a polarization of 9.8 $\mu$C/cm$^2$. 
Compared with the $Pbam$ phase at the same pressure, it shows a contraction in volume of 0.2\% and a lower mechanical enthalpy by about 3.6 meV/f.u., indicating enhanced stability.
Moreover, its phonon dispersion displays no imaginary frequencies (Fig. S4 \cite{SM}), confirming its dynamical stability under pressure.

The stabilization of the $Pmc2_1$ phase is not exclusive to hydrostatic compression.
We further find that suitably oriented biaxial compressive strain can also stabilize this phase (Supplementary Section S4 \cite{SM}).
These results demonstrate that the additional in-phase tilts required can be stabilized through distinct routes, pointing to a more general tendency toward the $Pmc2_1$ phase under lattice contraction. 

\begin{figure}[t!]
\centering
\includegraphics[scale=0.42]{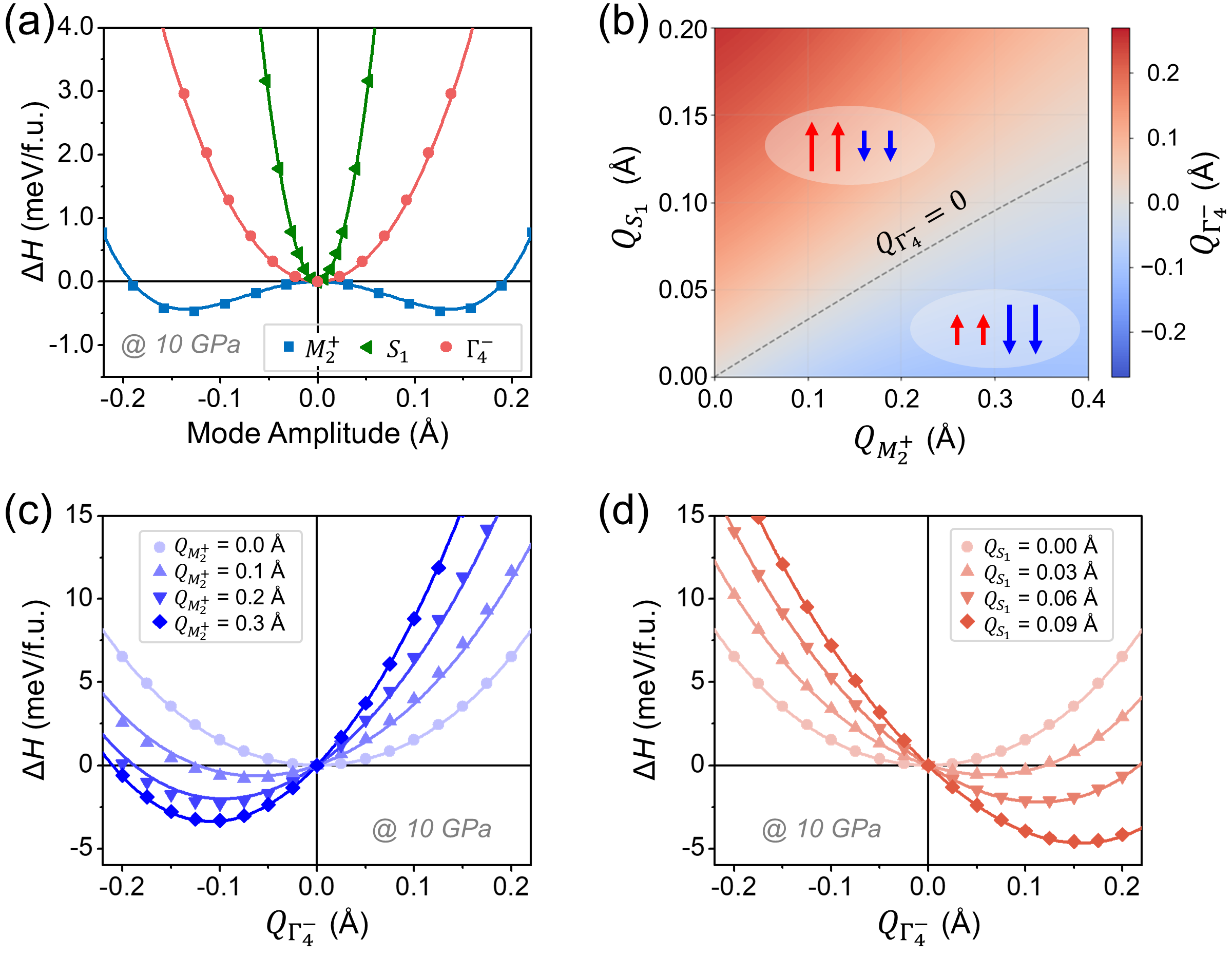}
\caption{Mode couplings in the $Pmc2_1$ phase at 10 GPa.
(a) Enthalpy gain from $Pbam$ as a function of $M_2^+$, $S_1$, and $\Gamma_4^-$ mode amplitudes.
(b) Relaxed $\Gamma_4^-$ amplitude at different $M_2^+$ and $S_1$ amplitude, illustrating a change in the sign of the equilibrium polarization across the $Q_{\Gamma_4^-}=0$ line.
(c, d) Enthalpy gain as a function of $\Gamma_4^-$ amplitude at several fixed amplitudes of (c) $M_2^+$ and (d) $S_1$ modes. 
Symbols represent first-principles results, while curves and contour map are derived from the fitted Landau model.
The $R_5^-$, $S_2$, and $\Sigma_2$ modes are fixed at their amplitudes in the $Pmc2_1$ phase.
Lattice parameters are relaxed under 10 GPa.
}
\label{fig:couplings}
\end{figure}

To substantiate the symmetry-based picture of tilt-driven FiE, we perform a group theoretical analysis of lattice-distortion couplings.
We focus on the six most dominant modes in the $Pmc2_1$ phase: the $R_5^-$, $S_2$, and $\Sigma_2$ modes inherited from the $Pbam$ structure, as well as the emergent $M_2^+$, $S_1$, and $\Gamma_4^-$ modes (Figs. S1 and S5 \cite{SM}).
Among these, the $\Gamma_4^-$ mode is the sole polar mode responsible for the polarization, whereas all other modes are nonpolar. 
The uncompensated ``$\uparrow \uparrow \shortdownarrow \shortdownarrow$'' configuration can be understood as the superposition of $\Sigma_2$ (``$\uparrow \uparrow \downarrow \downarrow$'') and $\Gamma_4^-$ (``$\uparrow \uparrow \uparrow \uparrow$'').
In addition, the $S_1$ mode, which can be interpreted as modulated tilt, is also included in our analysis due to its non-negligible amplitude and significant influence on the polarization.

Our analysis confirms that the $Pbam$-to-$Pmc2_1$ transition is primarily driven by the $M_2^+$ tilt mode, as it is the only mode that develops a double-well potential in the presence of $R_5^-$, $S_2$, and $\Sigma_2$ under 10 GPa [Fig. \ref{fig:couplings}(a)].
Once the $M_2^+$ mode condenses in the $Pbam$ structure, it then activates the $S_1$ and $\Gamma_4^-$ modes through symmetry-allowed couplings with pre-existing $Pbam$ distortions.

Our analysis reveals two distinct coupling pathways by which the $M_2^+$ tilt induces polarization ($\Gamma_4^-$).
The first is a direct pathway, where $M_2^+$ couples directly with the inherited $Pbam$ distortions ($\Sigma_2$ and $S_2$) to induce $\Gamma_4^-$:
\begin{equation}
\begin{split}
 Q_{\Sigma_2}^2 Q_{M_2^+} Q_{\Gamma_4^-}, \ 
 Q_{S_2}^2 & Q_{M_2^+} Q_{\Gamma_4^-}.
\end{split}
\label{eq:eq1}
\end{equation}
The second is an indirect pathway mediated by the $S_1$ mode. 
Here, $M_2^+$ first induces $S_1$ via
\begin{equation}
\begin{split}
 Q_{R_5^-} Q_{\Sigma_2} Q_{M_2^+} Q_{S_1}, \
 Q_{S_2} Q_{M_2^+} Q_{S_1},
\end{split}
\label{eq:eq2}
\end{equation}
and subsequently, $S_1$ contributes to the induction of $\Gamma_4^-$ through
\begin{equation}
\begin{split}
 Q_{R_5^-} Q_{\Sigma_2} Q_{S_1} Q_{\Gamma_4^-}, \
 Q_{S_2} Q_{S_1} Q_{\Gamma_4^-}, {\rm and} \
 Q_{M_2^+} Q_{S_1}^2 Q_{\Gamma_4^-}.
\end{split}
\label{eq:eq3}
\end{equation}
Detailed Landau expansion and fitted coefficients are provided in Supplementary Section S5 \cite{SM}.
We have also confirmed that this dual-pathway scenario remains robust, regardless of whether the lattice parameters are relaxed at 10 GPa or fixed to those of the cubic structure or the $Pmc2_1$ phase at 10 GPa.

A striking consequence of this dual-pathway scenario is the sensitivity of the polarization direction to the relative amplitudes of the $M_2^+$ and $S_1$ modes [Fig. \ref{fig:couplings}(b)].
As illustrated in Figs. \ref{fig:couplings}(c, d), the two modes favor polarization in opposite directions, so that the net polarization is determined by their subtle competition. 
Remarkably, this competition enables the polarization to evolve continuously through zero and reverse sign within the $Pmc2_1$ phase simply by tuning the relative amplitudes of $M_2^+$ and $S_1$, without requiring a sign reversal of the modes themselves [Fig. \ref{fig:couplings}(b)].

\begin{figure}[t!]
\centering
\includegraphics[scale=0.45]{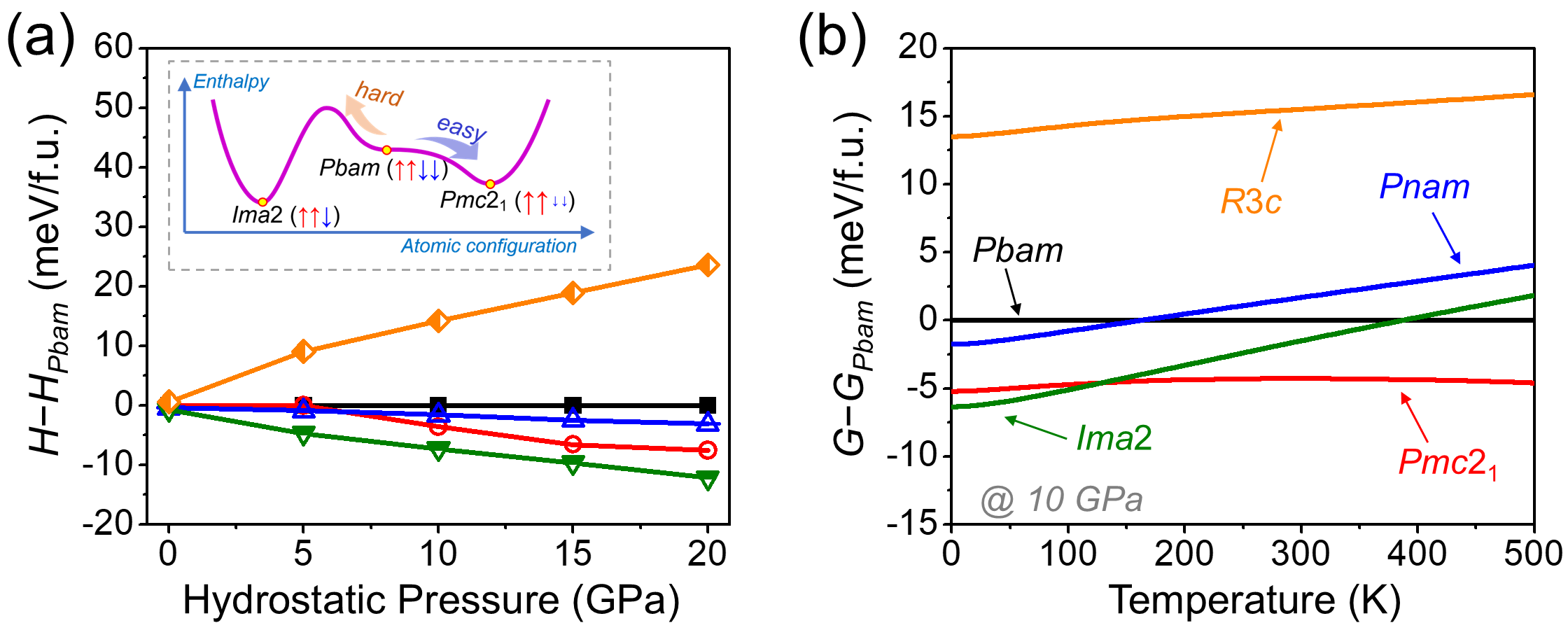}
\caption{Thermodynamic stability and kinetic accessibility of the competing phases.
(a) Enthalpy differences relative to $Pbam$ as a function of hydrostatic pressure. 
At 0 GPa, $Pmc2_1$ is unstable and relaxes to $Pbam$.
The inset illustrates energetic landscape under pressure, showing a barrier-free $Pbam$-to-$Pmc2_1$ transition and a kinetically hard $Pbam$-to-$Ima2$ pathway due to the barrier associated with reconstructing the polarization periodicity. 
(b) Gibbs free-energy differences relative to $Pbam$ at 10 GPa as a function of temperature, calculated within the quasi-harmonic approximation.
The same color scheme is used in both panels: black ($Pbam$), red ($Pmc2_1$), blue ($Pnam$), green ($Ima2$), and orange ($R3c$). 
}
\label{fig:enthalpy}
\end{figure}

We then compare the stability of the $Pmc2_1$ phase against other competing low-energy phases, including the conventional AFE $Pbam$ phase, the field-induced FE $R3c$ phase \cite{PhysRevB110054109, RN135}, the 80-atom AFE $Pnam$ phase \cite{RN77}, and the FiE $Ima2$ phase exhibiting ``$\uparrow \uparrow \downarrow$'' ordering \cite{RN90}.
As shown in Fig. \ref{fig:enthalpy}(a), although the $Pmc2_1$ structure relaxes back to $Pbam$ at 0 GPa, it is energetically more favorable than $Pbam$ above 5 GPa, and further stabilizes over $Pnam$ above 10 GPa.
However, the $Pmc2_1$ phase never becomes the ground state under hydrostatic pressure; instead, the FiE $Ima2$ phase has the lowest enthalpy throughout the pressure range considered.

Nevertheless, a slightly higher enthalpy than that of $Ima2$ does not preclude the experimental realization of the $Pmc2_1$ phase, especially when kinetic factors are taken into account.
In PZO, although the $Ima2$ phase is predicted to be the ground state, the $Pbam$ phase is the most commonly observed in experiments.
This can be understood as a consequence of kinetic hindrance: the transition from $Pbam$ to $Ima2$ requires a reconstruction of the polarization periodicity, from fourfold in $Pbam$ to threefold in $Ima2$.
Such a reconstruction inevitably involves a substantial energy barrier, making the $Pbam$-to-$Ima2$ transition kinetically unfavorable \cite{RN90, PhysRevB110054109}.
In contrast, the $Pmc2_1$ phase is structurally closer to $Pbam$, sharing the same fourfold periodicity.
Moreover, the two phases are symmetrically connected via a direct group-subgroup relation (Supplementary Section S6 \cite{SM}).
Consequently, once the thermodynamic condition (e.g., pressure) favors $Pmc2_1$ over $Pbam$, the transition can proceed readily without requiring a complex rearrangement of the polarization pattern [inset of Fig. \ref{fig:enthalpy}(a)].
Thus, the $Pmc2_1$ phase remains a kinetically accessible metastable state that can effectively compete with the global ground state.

We further evaluate the finite-temperature stability of these phases within quasi-harmonic approximation (QHA) under pressure.
Fig. \ref{fig:enthalpy}(b) presents the temperature dependence of Gibbs free-energy at 10 GPa, accounting for the phonon contribution to entropy and thermal expansion effects.
For the dynamically unstable $Pbam$ phase (Fig. S11 \cite{SM}), imaginary phonon modes were excluded from the vibrational entropy calculation; although approximate, this approach captures the essential trends \cite{RN90}.
Our results indicate that while $Ima2$ is the most stable phase at low temperatures ($<120$ K), the $Pmc2_1$ phase becomes thermodynamically more favorable around room temperature.
Moreover, the $Pmc2_1$ consistently shows lower Gibbs free-energy than both $Pbam$ and $Pnam$ over the entire temperature range (from 0 to 500 K), suggesting that a transition from $Pmc2_1$ back to $Pbam$ upon heating is unlikely. 
Instead, the $Pmc2_1$ phase may transform directly into the high-temperature cubic phase upon heating without passing through the $Pbam$ phase as an intermediate.
Similar trends are observed when employing the harmonic approximation without thermal expansion (Fig. S15 \cite{SM}).
Overall, these results suggest that the $Pmc2_1$ phase could be experimentally accessible near room temperature under appropriate lattice-contraction conditions.

\begin{figure}[t!]
\centering
\includegraphics[scale=0.34]{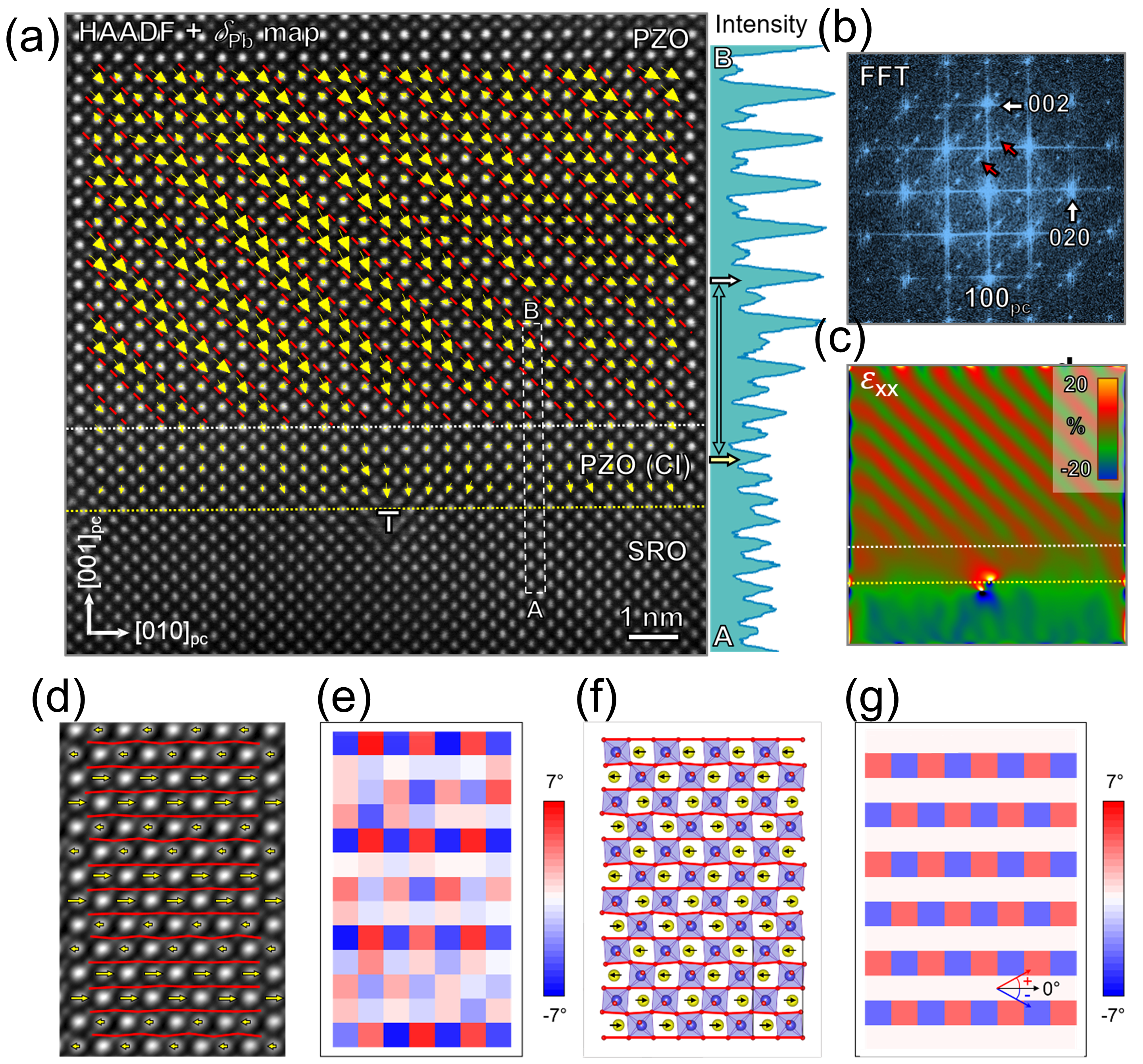}
\caption{Experimental evidence of $Pmc2_1$-like FiE structure in thin PZO thin films.
(a) STEM-HAADF image near the PZO/SRO interface, overlaid with a Pb displacement map (yellow arrows).
The intensity profile from ``A'' to ``B'' is used to identify the position of PZO/SRO interface.
A chemically intermixed (CI) PZO layer is observed above the interface, followed by a FiE region exhibiting uncompensated ``$\uparrow \uparrow \shortdownarrow \shortdownarrow$'' pattern.
(b) FFT pattern of the FiE region. 
(c) GPA in-plane strain map showing periodic strain modulation.
(d) iDPC image of the FiE region, in which red lines trace the rippling of  horizontal oxygen chains.
(e) Oxygen-chain rippling angles extracted from (d).
(f) Schematic crystal structure of the $Pbam$ AFE phase and (g) the corresponding map of oxygen-chain rippling for comparison.
}
\label{fig:experiment}
\end{figure}

Encouragingly, the $Pmc2_1$-like FiE structures are indeed observed experimentally in PZO, including both thin films and single crystals (refer to Supplementary Section S1.2 \cite{SM} for experimental details).
Fig. \ref{fig:experiment}(a) displays a high-resolution STEM-HAADF image acquired from a 220 nm PZO thin film grown on SrTiO$_3$ (STO) (001) substrate with SrRuO$_3$ (SRO) bottom electrode (see Fig. S16 for the film characterizations \cite{SM}).
Overlaid with a Pb displacement map (yellow arrows), this image reveals a distinct FiE region above the PZO/SRO interface, where the Pb displacements adopt an uncompensated ``$\uparrow \uparrow \shortdownarrow \shortdownarrow$'' ordering, consistent with the predicted $Pmc2_1$ phase.
The fast Fourier transform pattern [Fig. \ref{fig:experiment}(b)] and in-plane strain map [Fig. \ref{fig:experiment}(c)] confirm a fourfold modulation period.
In this FiE region, the principal Pb displacement component points toward the lower right, while the secondary component points toward the upper left, resulting in a net polarization directed towards the PZO/SRO interface.
Although this downward polarization orientation is more frequently observed near the PZO/SRO interface, suggesting a potential influence of the built-in interfacial electric field, regions with upward-oriented principal Pb displacements are also detected (Fig. S17 \cite{SM}).
Similar uncompensated Pb displacement patterns have been reported previously in PZO \cite{RN1275}; however, this feature alone is not sufficient to establish the proposed $Pmc2_1$ phase, motivating further examination of the oxygen sublattice.

To this end, we then focus on the oxygen octahedral tilts ($M_2^+$ mode), which is another defining structural feature of the $Pmc2_1$ phase that enables the tilt-driven mechanism. 
To probe this distortion, we employ the STEM integrated differential phase-contrast (iDPC) technique \cite{LAZIC2016265, LiuJAP2021} to quantify the ripples of oxygen chains.
As shown in Figs. \ref{fig:experiment}(d, e), the oxygen chains exhibit significant rippling, with the rippling amplitude between two large Pb displacements being distinctly different from that between two small Pb displacements.
Such asymmetric oxygen-chain rippling is incompatible with the $Pbam$ symmetry [Figs. \ref{fig:experiment}(f, g)], but it is a signature of the $M_2^+$ tilt mode in the $Pmc2_1$ phase.
The observation of this oxygen sublattice distortion, therefore, provides definitive structural evidence supporting the presence of the $Pmc2_1$-like FiE state.

Beyond the PZO/SRO interface, the $Pmc2_1$-like structures are also observed in the interior of the PZO thin film (Fig. S18 \cite{SM}), and in PZO single crystal (Figs. S19 and S20 \cite{SM}). 
The single-crystal specimen was annealed prior to TEM observation and displays dot-like contrasts in TEM images (Fig. S19 \cite{SM}), likely associated with Pb vacancies \cite{WeiADFM}.
Second-harmonic generation imaging reveals polar clusters, while HAADF imaging confirms the presence of uncompensated ``$\uparrow \uparrow \shortdownarrow \shortdownarrow$'' regions (Fig. S20 \cite{SM}).
Given that Pb vacancies are known to induce lattice contraction \cite{RN1190}, this suggests that point defects, such as Pb vacancies, may play a role in promoting the local $Pmc2_1$-like FiE structures in PZO.
In summary, the combined STEM-HAADF and iDPC measurements provide compelling experimental evidence for the existence of $Pmc2_1$-like FiE structures in both thin films and bulk PZO, thereby strongly supporting the theoretical mechanism proposed in this work.

The above results, combining first-principles calculations and experimental observations, establish a symmetry-governed mechanism for FiE in PZO.
Central to this mechanism is the octahedral tilt $M_2^+$ mode.
Condensing this mode into the $Pbam$ structure alone is sufficient to achieve the $Pmc2_1$ symmetry, while all other distortions, including polarization, emerge as secondary symmetry-allowed responses \footnote{This can also be understood as a consequence of the $\Gamma$-instability of the $Pbam$ phase which exhibits mixed character when decomposed into the irreducible representations of the cubic reference phase; see Fig. S12 for details \cite{SM}.}.
The stabilization of the $Pmc2_1$ phase is naturally promoted by lattice contraction, which geometrically favors tilts in the corner-sharing network.
In practical systems, such contraction can be achieved through microscopic sources of strain, such as epitaxial constraints or point defects (e.g., Pb vacancies), which can effectively reduce the external pressure required.

The identification of $Pmc2_1$ phase sheds light on several long-standing issues in PZO.
As a polar structure realized experimentally, it is likely to contribute to the observed unclosed double $P$-$E$ loop and the low-temperature polarity \cite{RN708}.
In addition, the in-phase octahedral tilts, although commonly seen in many perovskites, have not yet been unambiguously associated with a specific phase in PZO.
Experimental $M$-point diffraction features observed in the intermediate state near the antiferroelectric transition have been suggested to be related to such tilts, but the structural details remain elusive \cite{RN144, RN687}.
While the $Pmc2_1$ structure itself is unlikely to represent this intermediate state, it nevertheless provides a clear example of a low-energy phase in PZO with in-phase octahedral tilts.
Moreover, the couplings by which $M$-point tilts induce polarization also imply a reciprocal effect, whereby electric fields (including local fields arising from polarization fluctuations or charged defects) could, in turn, excite the $M$-point tilts through the same anharmonic couplings. 

Beyond PZO, the same $Pmc2_1$ phase can also be stabilized in other perovskites, such as PbSnO$_3$, where notably no external pressure is required (Fig. S21 \cite{SM}), indicating that the mechanism is not unique to a single compound.
More generally, the mode couplings underlying the tilt-driven FiE mechanism are broadly applicable.
The essential ingredients are the polarization modulation ($\Sigma_2$ mode, not limited to the fourfold periodic ``$\uparrow \uparrow \downarrow \downarrow$'') and in-phase octahedral tilts ($M_2^+$ mode).
Although both modes are nonpolar, their coexistence breaks inversion symmetry and induces the polar $\Gamma_4^-$ mode.
A related manifestation is found in 180$^\circ$ domain walls in the ferroelectric $Q$-phase of NaNbO$_3$ (Fig. S22 \cite{SM}).
There, combining in-phase tilts inherent in the $Q$-phase \cite{RN1255, RN600} with the effective polarization modulation created by the 180$^\circ$ domain walls results in unexpected asymmetry between oppositely polarized domains.
This example underscores that the tilt-driven FiE framework extends beyond PZO and the specific $Pmc2_1$ phase and is rooted in a more general symmetry principle.

The $Pbam$-to-$Pmc2_1$ transition under pressure in PZO also provides an example of pressure-induced polarization.
In contrast to proper ferroelectrics such as BaTiO$_3$, where pressure typically suppresses polarization, the polarization in $Pmc2_1$ originates from tilt-mediated improper couplings and emerges upon the appearance of the additional in-phase tilt under compression.
Meanwhile, this mechanism of generating polarization in $Pmc2_1$ is also more intricate than the conventional hybrid-improper ferroelectricity observed in superlattices and layered perovskites, where the polarization arises due to a single trilinear coupling term \cite{Bousquet2008, Benedek2011, RN23, RN597}.
In $Pmc2_1$, the relevant coupling terms are not trilinear, but instead the less common quadratic-linear-linear and quadrilinear terms [Eqs.~(\ref{eq:eq1}-\ref{eq:eq3})], previously seen in a few examples of brownmillerite and perovskite compounds \cite{RN1276, RN1277}.
Moreover, the polarization in $Pmc2_1$ is not governed by one single dominant coupling, but is rather determined by the subtle competition of multiple improper couplings.
This behavior represents a distinct form of ``competitive'' improper mechanism.

Finally, the tilt-driven FiE mechanism has implications for materials design. 
For energy-storage applications, where $P$-$E$ loops with vanishing remanent polarization are desired \cite{RN526}, avoiding the conditions that stabilize the $Pmc2_1$ phase is critical.
Conversely, emerging applications requiring multiple polarization states, such as multi-level ferroelectric memories or neuromorphic architectures \cite{RN1258, RN1259}, may exploit this $Pmc2_1$ phase.
Ultimately, the discovery of the $Pmc2_1$ phase highlights the structural richness and complexity of the potential landscape of PZO, suggesting that additional unconventional phases may remain to be discovered and could offer unexpected functional opportunities.

\vspace{3mm}
\textit{Acknowledgments.}
We thank Prof. Zijian Hong and Dr. Xiangwei Guo from Zhejiang University, and Prof. Hao-Cheng Thong from Beijing University of Posts and Telecommunications for their helpful discussions.
We also thank Dr. Alexander Block and Prof. Klaas-Jan Tielrooij for their assistance with the SHG experiment.
This work is supported by the European Union’s Horizon 2020 research and innovation programme under Grant Agreement No. 964931 (TSAR) and by F.R.S.-FNRS Belgium under PDR Grants No. T.0107.20 (PROMOSPAN) and No. T.0128.25 (TOPOTEX).
H.Z. acknowledges support from the Fundamental Research Funds for the Central Universities of China (Grant No. 104972026KFYjc0115).
Y.L. acknowledges support from the BIST Postdoctoral Fellowship Programme (PROBIST) funded by the European Union’s Horizon 2020 research and innovation programme under Marie Skłodowska-Curie Grant Agreement No. 754510.
S.G. acknowledges support from the PREBIST Cofund grant. This project received funding from the European Union’s Horizon 2020 research and innovation programme under Marie Skłodowska-Curie Grant Agreement No. 754558.
B.X. acknowledges support from the National Natural Science Foundation of China (Grant No. 12574101).
The authors acknowledge the use of the CECI supercomputer facilities funded by the F.R.S.-FNRS (Grant No. 2.5020.1) and the Tier-1 supercomputer of the Fédération Wallonie-Bruxelles funded by the Walloon Region (Grant No. 1117545).
The authors are grateful for the scientific and technical support from Sydney Microscopy and Microanalysis (SMM). SMM is a Core Research Facility of the University of Sydney and a foundational node of Microscopy Australia (ROR: 042mm0k03), which is supported by the Australian Government’s National Collaborative Research Infrastructure Scheme.

\bibliography{reference}
\bibliographystyle{unsrt}

\end{document}